\documentclass[fleqn,10pt]{wlscirep}
\usepackage[utf8]{inputenc}
\usepackage[T1]{fontenc}

\usepackage{cite}
\usepackage{amsmath,amssymb,amsfonts}
\usepackage{graphicx}
\usepackage{longtable}
\usepackage{textcomp}
\usepackage{gensymb}
\usepackage{comment}
\usepackage{url}
\def\BibTeX{{\rm B\kern-.05em{\sc i\kern-.025em b}\kern-.08em
    T\kern-.1667em\lower.7ex\hbox{E}\kern-.125emX}}

\usepackage{acronym}
\usepackage{algorithm}
\usepackage{algpseudocode} 
\usepackage{caption}
\newacro{AI}[AI]{Artificial Intelligence}
\newacro{APC}[APC]{Automated Passenger Counter}
\newacro{API}[API]{Application Programming Interface}
\newacro{AR}[AR]{Augmented Reality}
\newacro{DLT}[DLT]{Distributed Ledger Technology}
\newacro{DoS}[DoS]{Denial of Service}
\newacro{HMD}[HMD]{Head-Mounted Display}
\newacro{QoS}[QoS]{Quality of Service}
\newacro{XR}[XR]{Extended Reality}
\newacro{ITS}[ITS]{Intelligent Transport System}
\newacro{REST}[REST]{REpresentational State Transfer}
\newacro{VR}[VR]{Virtual Reality}
\newacro{IPFS}[IPFS]{InterPlanetary File System}
\newacro{IoT}[IoT]{Internet of Things}
\newacro{MQTT}[MQTT]{Message Queue Telemetry Transport}
\newacro{UWP}[UWP]{Universal Windows Platform}
\newacro{MRTK}[MRTK]{Mixed Reality Toolkit}

\title{A Comprehensive Survey on Green Blockchain: Developing the Next Generation of Energy Efficient and Sustainable Blockchain Systems}

\author[1,2]{Tiago M. Fernández-Caramés}
\author[1,2,*]{Paula Fraga-Lamas}
\affil[1]{Department of Computer Engineering, Faculty of Computer Science, Universidade da Coru\~na, \mbox{15071 A Coru\~na, Spain}}
\affil[2]{Centro de Investigaci\'on CITIC, Universidade da Coru\~na, \mbox{15071 A Coru\~na, Spain}}

\affil[*]{tiago.fernandez@udc.es, paula.fraga@udc.es}

\begin{abstract}
Although Blockchain has been successfully used in many different fields and applications, it has been traditionally regarded as an energy-intensive technology, essentially due to the past use of inefficient consensus algorithms that prioritized security over sustainability.
However, in the last years, thanks to the significant progress made on key blockchain components, their energy consumption can be decreased noticeably.
To achieve this objective, this article analyzes the main components of blockchains and explores strategies to reduce their energy consumption. In this way, this article delves into each component of a blockchain system, including consensus mechanisms, network architecture, data storage and validation, smart contract execution, mining and block creation, and outlines specific strategies to decrease their energy consumption.
For such a purpose, consensus mechanisms are compared, recommendations for reducing network communications energy consumption are provided, techniques for data storage and validation are suggested and diverse optimizations are proposed both for software and hardware components.
Moreover, the main challenges and limitations of reducing power consumption in blockchain systems are analyzed.
As a consequence, this article provides a guideline for the future researchers and developers who aim to develop the next generation of Green Blockchain solutions.
\end{abstract}
\begin{document}

\flushbottom
\maketitle
%
%


\section{Introduction}
\label{sec:introduction}

In 2008 the proposal of Bitcoin showed that it was possible to implement a distributed cryptocurrency without requiring trusted third parties \cite{Nakamoto2008}. That was possible thanks to joining together several previous concepts like Proof-of-Work (PoW) \cite{Jakobson1999}, hash functions \cite{Wang2020}, distributed timestamping \cite{Maasias1999} and Merkle trees \cite{Merkle1987}. Such a technology combination resulted in the creation of the Blockchain technology, which has been employed in multiple fields and applications \cite{Algha2024}.

Blockchain systems implement a type of \ac{DLT} that, in the particular case of those based on PoW consensus mechanisms, have been known for their significant energy consumption \cite{Sedlmeir2020}, since they require that part of the participants solve complex mathematical puzzles to validate transactions and to secure the network \cite{Baniata2023}. 
This is due to Sybil attacks \cite{Douceur2002}, which pose a critical problem for DLT systems, and which require an attacker to create multiple fake identities to take control of the decisions on the blockchain. Traditionally, it has been considered that, to control a blockchain, 51\% of the computing power was necessary (thus performing what is called a `51\% attack'), but researchers have demonstrated in the last years that, in a large blockchain like Bitcoin, it is sufficient with a percentage of 32\% \cite{Zhang2019}.
 To prevent such attacks, permissioned blockchains can control the access to rogue participants \cite{Sedlmeir2022}, but, in permissionless networks, where participant access is not restricted, complex mechanisms like PoW consensus protocols are needed.

The problem is that PoW consensus protocols involve high computational power and, consequently, substantial energy consumption. Bitcoin \cite{Nakamoto2008}, the most well-known blockchain, has drawn attention due to its high energy footprint, with estimates of its energy consumption surpassing that of some countries \cite{DeVries2018}. In fact, a single Bitcoin PoW-based transaction requires the energy demanded by, for instance, an average German household for weeks or months \cite{Sedlmeir2020}. That is the reason why certain countries of the European Union asked for the ban of energy-intensive blockchain activities \cite{Kshetri2022}. This high energy consumption has also raised concerns about the environmental impact and the sustainability of blockchain technology \cite{Truby2018}. Some researchers went further and estimated that, if a PoW-based DLT like Bitcoin was used at a global scale, the associated emissions would lead to a 2\degree{C} temperature increase in the coming decades \cite{Mora2018} (such an estimation has been already criticized and debunked by other researchers \cite{Dittmar2019,Houy2019,Masanet2019}, which concluded that PoW blockchains are not a threat to the climate \cite{Sedlmeir2020}).

All the previously mentioned assessments on Bitcoin, unfortunately, still dominate people's perceptions on what a blockchain is and how much energy is necessary to make it work, frequently encountering claims that blockchain energy consumption is problematic \cite{Sedlmeir2020}, thus neglecting the major energy-consumption improvements achieved since when Bitcoin was conceived (in 2008). In fact, the original Bitcoin blockchain was referred as `Blockchain 1.0', while the evolution towards smart contracts was named `Blockchain 2.0' and the development of applications beyond cryptocurrencies and finance (especially in areas like government, health or art) as `Blockchain 3.0' \cite{Swan2015}.

Therefore, the evolution of blockchain technologies and their improvements are still on-going, but it is still necessary to decrease farther their energy consumption due to their impact on several fields:

\begin{itemize}
    \item Environmental Impact. The energy consumption of blockchain systems contributes to carbon emissions and exacerbates climate change \cite{Truby2018}. As societies strive to transition to a more sustainable and low-carbon future, it is essential to address the energy consumption of blockchain technology \cite{Sedlmeir2020}.

    \item Energy Efficiency. Improving energy efficiency in blockchain systems can lead to cost savings for users and operators \cite{Platt2022}. Lower energy consumption means reduced operational expenses, making blockchain technology more economically viable and attractive to adopt \cite{Sedlmeir2020}.

    \item Scalability and Adoption. High energy consumption limits the scalability of blockchain systems, making it challenging to handle a large number of transactions \cite{Sanka2021}. By reducing energy consumption, blockchain systems can become more scalable, enabling wider adoption and integration into various industries \cite{Liu2020}.

    \item Social Responsibility. Emphasizing energy efficiency aligns with the broader principles of corporate social responsibility and ethical innovation \cite{Harmon2011}. Blockchain technology can be used for social good and positive impact if it is designed and implemented with environmental considerations in mind \cite{Fraga2020}.

\end{itemize}

This article analyzes current blockchain technologies together with their challenges and limitations in terms of energy efficiency in order to create Green Blockchains. In particular, the following are the main contributions of this article, which, as of writing, have not been found together in the literature:

\begin{itemize}
    \item It analyzes the essential components of a blockchain in order to determine the main software and hardware contributors to energy consumption.
    
    \item It explores the most relevant energy-saving strategies for each component of a blockchain system, including consensus mechanisms, network architecture, data storage and validation, smart contract execution and mining/block creation.



    \item The main challenges and limitations for implementing energy-efficient blockchains are discussed, considering the trade-offs between energy efficiency and security, the problem of estimating energy consumption in blockchain, the regulatory and governance aspects that impact energy efficiency and sustainability, or the scalability/performance implications of energy consumption reductions.

\end{itemize}

The rest of the article is structured as follows. Section \ref{sec:overview} is dedicated to describing the main characteristics and components of blockchain systems. Section \ref{sec:components} analyzes the energy-intensive components of a blockchain, while Section \ref{sec:strategies} suggests multiple strategies to reduce the power consumption of such components. Next, Section \ref{sec:challenges} describes the main challenges and limitations of implementing energy-efficient blockchain-based solutions. Finally, Section \ref{sec:conclusions} is devoted to conclusions.

\section{Main characteristics and components of a blockchain system}
\label{sec:overview}

In order to minimize blockchain energy consumption, it is first necessary to understand how a blockchain operates. Thus, the next subsections provide the definition of blockchain, including details on its essential software and hardware.

\subsection{Definition of blockchain}

First, it is essential to define what blockchain is: a distributed ledger that can store and verify transactions without relying on a central authority or intermediary. Such a technology can enable various applications, such as cryptocurrencies \cite{Primecoin}, smart contracts \cite{Khan2021}, supply chain management \cite{Dud2024} or digital identity \cite{Zeydan2024}.

 \begin{figure*}[!hbt]
     \centering
     \includegraphics[scale=0.3]{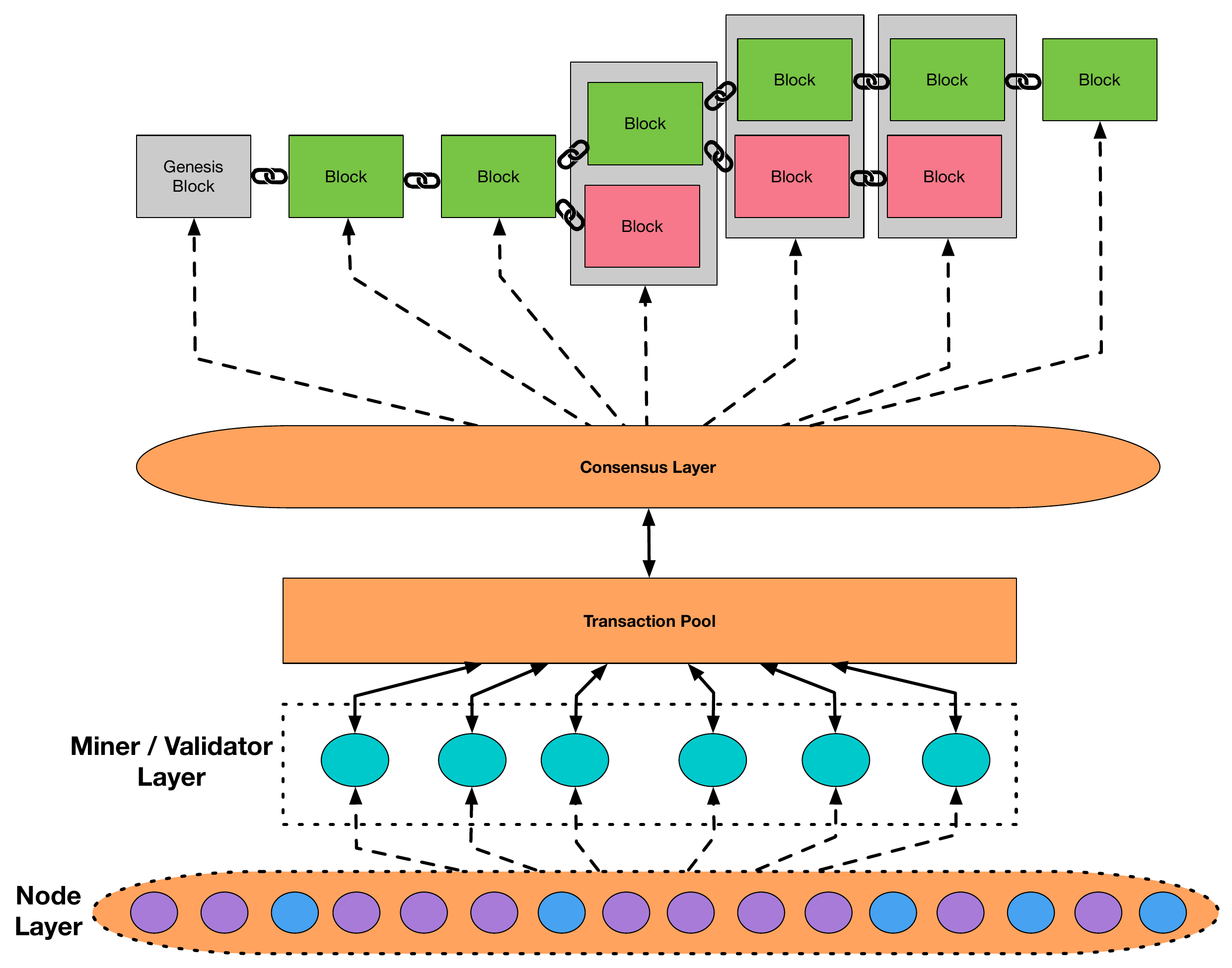}
     \caption{Main components of a blockchain.}
     \label{figure:components}
\end{figure*}

 The main features of blockchain technology are:
    
    \begin{itemize} 
        \item Decentralization: blockchains rely on a network of nodes that can validate and update the ledger without a central point of control or failure. 
        
        \item Immutability: blockchain technologies are designed to resist  tampering and unauthorized modifications carried out once a transaction is recorded and verified by the network. 
        
        \item Transparency: blockchains allow anyone with access to the ledger to view and to verify the transactions and their history. 
        
        \item Security: blockchains make use of cryptographic techniques such as digital signatures and hash functions, to protect the data and transactions from unauthorized access or modification. 
        
    \end{itemize}

In order to provide such features, a blockchain system needs to be composed of several essential elements, which are illustrated in Figure \ref{figure:components} and are described in the following subsections.

\subsection{Essential blockchain software}

A blockchain is composed by software that implements the following main functionality:

\begin{itemize}
    \item Distributed ledger. In terms of software, a blockchain is implemented as a distributed ledger that chronologically and immutably records all transactions and data \cite{Nakamoto2008}. Specifically, it contains:

        \begin{itemize}

            \item A ledger. It is the actual data stored by the system as a database or record of all transactions that have been validated and added to the blockchain. The ledger is shared and synchronized among all nodes in the network (i.e., the ones on the Node Layer in Figure \ref{figure:components}).
            
            \item Blocks. The block is the unit of storage of a blockchain and contains a batch of transactions together with other metadata, like a timestamp, a nonce, a hash of the previous block and a hash of the current block. As it can be observed in Figure \ref{figure:components}, blocks are linked together to form a chain of transactions. There are three types of blocks in Figure \ref{figure:components}:
            \begin{itemize}
                \item Genesis block. It is the first block, and thus the one that initiates the blockchain.

                \item Validated blocks. They are represented in green in Figure \ref{figure:components}. They are blocks that have been approved by the blockchain peers to be included in the blockchain.

                \item Non-validated blocks (in red in Figure \ref{figure:components}). They are also called orphan blocks and represent blocks that, due to multiple reasons, have not been included in the actual blockchain. For instance, such blocks may occur when some nodes perform the consensus procedure faster than others (i.e., the block that is approved the sooner is the one added to the blockchain). Moreover, non-validated blocks may be stored temporarily off-chain, so they cannot be added to the blockchain until they are propagated and validated by blockchain miners/validators.
            \end{itemize}

            \item Transactions. They represent anything of value, such as money, goods, services or data. Transactions are generated continuously on the blockchain and end up in a transaction pool where they are packed into blocks that are then approved by authorized miners/validators through a consensus mechanism.
       
        \end{itemize}

    \item Consensus Mechanism. The consensus mechanism defines how participants agree on the validity and order of transactions. Common consensus mechanisms include PoW, Proof-of-Stake (PoS), Delegated Proof-of-Stake (DPoS) and Practical Byzantine Fault Tolerance (PBFT) \cite{Zheng2017} (more detail on consensus mechanisms is included in Section \ref{sec:consensusMechanisms}).

    \item Network protocols. These protocols allow participants to communicate and coordinate in the blockchain system. The network usually functions as a peer-to-peer network where nodes keep copies of the blockchain and exchange information.
    
    \item Data Storage and validation. This component deals with the storage and validation of data on the blockchain. It includes methods to ensure the integrity and authenticity of transactions, such as cryptographic hashing and digital signatures \cite{Zysk2015}.
    
    \item Smart Contracts. Smart contracts are contracts that are executed with predefined rules and conditions encoded on the blockchain. They automate and enforce the execution of agreements, eliminating the need for intermediaries \cite{Szabo1996}.
    
    \item Mining. Mining is the process by which new transactions are validated, appended to the blockchain, and consensus is achieved. Traditionally, miners have used computational power to solve complex mathematical puzzles to secure the network and earn rewards \cite{Vranken2017}.

\end{itemize}

\subsection{Essential blockchain hardware}

Hardware nodes are the participants or entities that run the blockchain software and maintain the ledger. Nodes can be classified into different types depending on their role and function in the network (in Figure \ref{figure:components}, the Node Layer shows nodes in different colors to represent such different roles). The most common blockchain node roles are:

    \begin{itemize}
        \item Full nodes: a full node is a computing devices that participates in a blockchain by maintaining a complete copy of the transaction history and by verifying every transaction and block. Moreover, full nodes are able to communicate with other blockchain nodes to share information about new transactions and blocks, what helps to ensure that the blockchain remains decentralized and up-to-date across the entire network. For example, in the Bitcoin blockchain a full node stores the entire blockchain and validates transactions using the consensus mechanism rules. In the case of Ethereum, a full node stores all transactions and smart contract executions, validating them against the network rules. As it can be guessed,  a full node requires a significant amount of storage, processing power and bandwidth, but it is essential to provide a high level of security to the blockchain.

        \item Light nodes: they participate in the network without needing to store or process the entire blockchain. Instead, they only download and verify small portions of the blockchain data (typically just the block headers), which allow them to verify transactions without needing to store the full transaction history. As a consequence, light nodes require a lower amount of storage, processing power and bandwidth than full nodes, but light nodes rely on full nodes to provide them with the necessary data. Such a lower need for computational resources and energy makes them easier to be run on resource-constraint or battery-dependent devices. For instance, Bitcoin light nodes usually download block headers and validate transactions by requesting the necessary data from full nodes. In the case of Ethereum, light nodes download a small portion of the blockchain and rely completely on full nodes for the verification tasks.

        \item Miners or Validators. They are blockchain participants that make use of computational power to validate transactions and create new blocks. Thus, miners/validators are essential for maintaining the security, integrity and operation of the blockchain.
        In order to function as a validator node, a participant is required to have a computer that can communicate over the Internet or on an Intranet. Such a computer should have the capability of carrying out the necessary computations for verifying proposed transactions and performing other calculations as stipulated by the consensus protocol. The act of operating a validator node is voluntary, allowing participants to decide whether they want to participate in the blockchain in this role. It is important for validator nodes to stay active since the periods of activity, which are randomly determined, cannot be anticipated in advance. This latter fact involves a significant energy consumption that other types of participants do not have.

    \end{itemize}

In terms of hardware, to operate in a blockchain, every node requires at least to have computational hardware (e.g., based on a CPU, a GPU, an Application-Specific Integrated Circuit (ASICs), a Field-Programmable Gate Array (FPGA) or on a Complex Programmable Logic Device (CPLD)), memory (i.e., RAM and hard disks) and communication interfaces.

\section{Energy-intensive components of a blockchain}
\label{sec:components}

Blockchain systems consume energy for various purposes, such as for the validation of transactions, for creating blocks, for maintaining the ledger or for communicating with other nodes. 
Some components of a blockchain system tend to use more energy than others. For instance, consensus protocol energy consumption has been traditionally high but, unfortunately, many publications just focus on it \cite{Platt2022}, neglecting the contribution of other blockchain components. To avoid such a limitation, the following subsections analyze not only the impact of consensus protocols on the blockchain energy footprint, but also the other essential components that contribute to the overall consumption.

\subsection{Consensus Mechanism}
\label{sec:consensusMechanisms}

The consensus mechanism is one of the most energy-intensive components of blockchain systems, since, as it was previously mentioned, it determines how the participating nodes agree on the state of the ledger and prevent double-spending or malicious attacks \cite{Sedlmeir2020}. 

Different consensus mechanisms have different energy requirements and trade-offs. For example, PoW-based consensus mechanisms like the one used by Bitcoin and other cryptocurrencies, requires nodes that compete to solve complex mathematical puzzles, thus consuming a large amount of computational power and electricity \cite{Sinan2019}. Algorithmically, Bitcoin PoW mining is really simple \cite{Heinonen2022}, as it can be observed in Algorithm \ref{fig:BitcoinMiningPseudocode}.

\begin{algorithm}
    \caption{Bitcoin PoW mining algorithm.} 
    \label{fig:BitcoinMiningPseudocode}
    
    \begin{algorithmic}[1]
    	
    	\State $nonce \gets MIN$
        \While{$nonce < MAX$}
            \If{$sha256(sha256(block+nonce)) \le target$}
                \State \textbf{return} \, $nonce$
            \EndIf
            \State $nonce \gets nonce + 1$
        \EndWhile
    \end{algorithmic}
\end{algorithm}

Essentially, Bitcoin mining is a brute-force search for a value called `nonce' that, once added to a specific block header, the hash of such a header is lower or equal to a target value established by the blockchain network. For instance, when using SHA-256 as hash function, a target value can be the following 256-bit value (expressed in hexadecimal):

{\tiny $0x00000000000059e9054aad62105a259726801d5f494acbfcd40591c82f9b3136$}

Thus, a generated value will be lower than the target when its number of leading zeros is larger than the ones of the target. As a consequence, the higher the number of leading zeros, the more difficult is to find a nonce to meet the target condition and, therefore, more energy consumption will be dedicated to the search.

To avoid the energy inefficiencies of PoW, in the last years multiple alternative consensus mechanisms have been proposed, like:

\begin{itemize}
    \item Proof-of-Stake (PoS). PoS is a consensus mechanism used by Ethereum 2.0 and other blockchain platforms, which require nodes to stake a certain amount of tokens or cryptocurrency to participate in the validation process. Such a process consumes less energy than PoW-based consensus mechanisms, but it may introduce centralization or security risks. The first practical implementation of PoS is said to be Peercoin (in 2012) \cite{Peercoin}. The original PoS minted blocks in a similar way to PoW-based system (i.e., a mathematical puzzle needed to be solved), but relied on what is called coin age: how much time an amount of coins has been held by a node. Thus, the difficulty of the mathematical puzzle to be solved was assigned individually and was inversely proportional to the user coin age (i.e., the higher the coin age, the lower the difficulty of the mathematical puzzle). In the case of Ethereum, as of writing, a node should have at least 32 ETH and a computer connected to the Internet 24/7 to become a validator, although it is possible for those who do not own 32 ETH to participate in the validation through pooled staking \cite{EthereumPoS}.

    \item Delegated Proof-of-Stake (DPoS): it is a consensus mechanism designed for efficiency, speed and scalability by allowing token holders to vote for a small number of delegates (sometimes called `witnesses') who manage block production and network validation on behalf of the entire network \cite{Zhao2022}. Each token holder voting power is proportional to the number of tokens it holds. Examples of blockchains that make use of DPoS are EOSIO \cite{EOSIO} and TRON \cite{TRON}.

    \item Byzantine Fault Tolerance (BFT): it is a consensus protocol designed to ensure that a network can continue functioning correctly even if some participants behave maliciously or unpredictably. Its name derives from the Byzantine Generals Problem, a theoretical problem that illustrates the difficulties in achieving consensus when some participants may be unreliable or deceitful \cite{Lamport1982}. In a BFT-based system, all nodes (or a subset of validator nodes) exchange messages to verify that a proposed block or transaction follows the network rules: if enough nodes agree on the validity, the block is added to the blockchain.

    \item Practical Byzantine Fault Tolerance (PBFT). It is a variation of the BFT consensus protocol that operates in rounds where nodes (usually called replicas) agree on the next block. Thus, a leader node proposes a block and other nodes validate it. The process involves multiple rounds of voting until a consensus is reached. PBFT is used in permissioned blockchains (where only authorized participants can be validators) like Hyperledger Fabric \cite{Hyperledger}.

    \item Federated Byzantine Agreement (FBA). It is another variation of BFT designed for providing scalability security and low energy consumption in mind. The main difference with BFT is that it allows participants to choose who they trust in a flexible and decentralized manner \cite{Yoo2019}. Thus, each blockchain node selects a set of participants it trusts, known as a `quorum slice'. A quorum is formed when enough overlapping quorum slices agree on a decision. Therefore, if enough quorum slices overlap, the network can reach consensus as long as there is sufficient agreement within those slices. An example of implementation of FBA is the Stellar Consensus Protocol (SCP) \cite{Stellar}.

    \item Delegated Byzantine Fault Tolerance (DBFT). It is another variant of BFT where token holders vote to select a small group of trusted validators \cite{Zhan2022}. Such validators then use a BFT process to reach consensus, similar to when using DPoS, but with stronger guarantees of fault tolerance. An example of use of DBFT is NEO \cite{NEO}.

    \item Proof-of-Authority (PoA): it relies on a small group of pre-approved validators (called `authorities') that are responsible for validating transactions and creating new blocks. Unlike other consensus mechanisms, PoA operates on the basis of the identity and reputation of validators. Thus, the identities of the authorities are typically publicly known and they stake their reputation (rather than tokens or computational resources) on their honesty and performance. Governance in PoA-based systems is often managed by a central authority or consortium that selects and manages the validators. This introduces a degree of centralization, but it also allows for greater control and security in certain use cases, such as enterprise or consortium blockchains \cite{Yang2022}. In fact, some authors do not consider PoA-based permissioned blockchains as actual blockchains, since their behaviour differs significantly from the original blockchain concept (as defined by Satoshi Nakamoto), where a key requirement for implementing a blockchain was the complete lack of trust among the participants. Moreover, in many cases when a blockchain consortium is conformed among participants that trust each other, a blockchain is not efficient \cite{Fernandez2018}.
    In any case, PoA avoids the need for energy-intensive mining and consensus is achieved quickly, since the validators are few and known, and they cooperate rather than compete.

    \item Proof-of-Importance (PoI): it is designed to reward participants based on their overall contribution to the network. In PoI each node is assigned an importance score, which is then used to determine the likelihood of being chosen to validate blocks and earn rewards. Such a score is calculated based on several factors, like the amount of held stake, transaction activity (the more frequent transactions, the higher the importance score) or the active participation in maintaining the network. The nodes with higher importance have a better chance of being selected as harvesters (block validators) and, as a consequence, of receiving block rewards and transaction fees. In addition, PoI supports a feature called `delegated harvesting', which allows users to delegate their importance score to another trusted node, which can harvest blocks on their behalf. However, it must be noted that the calculation of the importance score can become complex and that there is a risk of centralization (especially in small networks), since large token holders can also engage heavily in transactions, so they can end up dominating the network.

    \item Proof-of-Burn (PoB): this consensus mechanism requires blockchain participants to `burn' tokens to demonstrate their commitment to the network. The process of `burning' tokens involves sending them to an address from which they can never be retrieved or used again, effectively removing them from circulation. Thus, by `sacrificing' tokens, participants earn the right to mine blocks or validate transactions, depending on the specific implementation.

    \item Proof-of-Capacity (PoC) or Proof-of-Space (PoSp): the participants allocate disk space (capacity) to mine new blocks \cite{Dziem2015}. Thus, PoC leverages available storage on a node hard drive to secure the network and validate transactions. The storage process involves precomputing and storing cryptographic solutions (called `plots') on the hard disk. Plots are usually hashes, typically derived from data that include the blockchain cryptographic hash function (e.g., SHA 256). Each hash represents a potential solution to a future block creation challenge.

    \item Proof-of-Luck (PoL) or Proof-of-Ellapsed-Time (PoET): it determines the next block producer based on random chance (luck) rather than computational power or token holdings. This randomness helps ensure fairness, while the reliance on secure hardware (Trusted Execution Environments, TEEs) ensures the integrity of the selection process. As a consequence, no intensive computation is required, but it is necessary to access hardware that supports TEEs (most modern processors do).

    \item Proof-of-Activity (PoAC): it is a hybrid consensus mechanism that combines aspects of both PoW and PoS to secure a blockchain network \cite{Bentov2014}. It was introduced as an attempt to address some of the energy inefficiencies and centralization risks associated with traditional PoW systems while also leveraging the fairness and decentralization of PoS. For such a purpose, PoAC operates in two phases:

    \begin{itemize}
        \item PoW phase: the initial phase is similar to other PoW-based consensus protocols (i.e., it makes use of miners to perform computational work to try to solve a cryptographic puzzle) and consists in mining a block that contains no transactions (it usually only includes the header information and miner identification information). 
        
        \item PoS phase: after a miner successfully mines the block in the PoW phase, a group of validators is selected randomly from a pool of stakeholders (the more tokens a participant holds, the higher the chances of being selected). Then, the selected validators are responsible for verifying and signing the block mined by the PoW miner.

    \end{itemize}

    \item Proof-of-Believability (PoBe): it is a consensus mechanism introduced by the IOST (Internet of Services Token) blockchain \cite{IOST} and that focuses on providing a high transaction volume together with a balance between scalability, decentralization and security. Similarly to PoI, PoBe evaluates a node’s credibility (or `believability') based on its contribution and behavior within the network, assigning a believability score that determines the likelihood of that node for being selected to validate and to produce the next block. Such a score depends on factors like reputation, past contributions to the network, held token and community trust (nodes can vote for other nodes or delegate on them). In this way, no intensive computational work is required, competition among nodes is reduced and the focus is essentially on reputation. However, it must be noted that reputation can bias the network, since the oldest and more established nodes can monopolize block production, especially in small networks.

    \item There are many other consensus mechanisms like Proof-of-Devotion \cite{Nebulas}, Proof-of-Bandwidth, Proof-of-Reputation \cite{Zhuang2019}, Proof-of-Download, Proof-of-Weight \cite{Bada2021}, Proof-of-Retrievability or Proof-of-Contribution \cite{Song2021}. An extensive survey of consensus mechanisms can be found in \cite{Xiao2020}.

    \end{itemize}

A summary of the previously mentioned consensus mechanisms is provided in Table \ref{tab:consensusMechanisms}. A detailed comparison on their energy efficiency is out of the scope of this article, but the interested reader can find further information in \cite{Bada2021,Bodkhe2020, Deval2024}.

\begin{longtable}{|p{4cm}|c|p{6cm}|l|}
\hline
\tiny \textbf{Consensus Mechanism} & \textbf{\tiny Energy Efficiency} & \textbf{Main Features} & \textbf{Examples of Blockchains} \\
\hline
Proof of Work (PoW) & Low & High computational effort; Miners solve cryptographic puzzles; Secure but energy-intensive & Bitcoin, Ethereum (pre-2.0) \\
\hline
Proof-of-Stake (PoS) & High & Validators are chosen based on the amount of staked tokens, reducing energy consumption & Ethereum 2.0, Cardano \\
\hline
Delegated Proof-of-Stake (DPoS) & High & Stakeholders select delegates that produce blocks; Fast and scalable & EOSIO, TRON \\
\hline
Byzantine Fault Tolerance (BFT) & High & Agreement among nodes in a decentralized system in the presence of malicious actors & Tendermint \\
\hline
Practical Byzantine Fault Tolerance (PBFT) & Medium & Optimized for fault tolerance and performance; Efficient for small networks & Hyperledger Fabric \\
\hline
Federated Byzantine Agreement (FBA) & High & Nodes select trusted validators to reach consensus & Stellar, Ripple \\
\hline
Delegated Byzantine Fault Tolerance (DBFT) & High & Delegates reach consensus on behalf of the network; Optimized for business use & NEO \\
\hline
Proof-of-Authority (PoA) & High & Validators are pre-approved; No mining; Suitable for private blockchains & VeChain, POA Network \\
\hline
Proof-of-Importance (PoI) & High & It considers factors like transaction activity, not just wealth, to determine validators & NEM \\
\hline
Proof-of-Burn (PoB) & Medium & Users burn coins to gain mining rights, simulating resource consumption without real energy costs & Counterparty \\
\hline
Proof-of-Capacity (PoC) & Medium & Validators use disk space to solve puzzles & Burstcoin \\
\hline
Proof-of-Luck (PoL) & High & Trusted execution environments generate random outcomes, eliminating need for energy-intensive mining & Intel SGX-based blockchains \\
\hline
Proof-of-Activity (PoAC) & Medium & PoW and PoS hybrid mechanism; Miners start the block creation and stakeholders finalize it & Decred \\
\hline
Proof-of-Believability (PoBe) & High & Nodes are selected based on reputation, past behavior and contributions to the network & IOST \\
\hline
Proof-of-Devotion (PoD) & High & Authority Masternodes selected based on commitment and community contribution & Nebulas \\
\hline
Proof-of-Reputation (PoR) & High & Validators are selected based on their reputation within the network; Often used in permissioned blockchains & GoChain \\
\hline
Proof-of-Download (PoDo) & High & Validators prove they have downloaded content to ensure data integrity and availability & File-sharing blockchains \\
\hline
Proof-of-Weight (PoWe) & High & Stake is weighted based on multiple factors (e.g., token ownership, resource contribution) & Algorand \\
\hline
Proof-of-Retrievability (PoRe) & High & Validators prove they can retrieve specific data efficiently; Used for storage networks & Filecoin \\
\hline
Proof-of-Contribution (PoCon) & High & Participants are rewarded based on their contributions to the network ecosystem and services & iExec \\
\hline
\caption{Comparison of some of the most popular consensus mechanisms and their energy efficiency.}
\label{tab:consensusMechanisms}
\end{longtable}

\subsection{Network architecture}

The network architecture is another energy-intensive component of blockchain systems, as it determines how the nodes are organized and connected in the network. Specifically, keeping a peer-to-peer network and spreading transaction information among nodes can need a lot of communications and energy resources, especially in large-scale networks.

Different network architectures have different energy implications and trade-offs. For example, public blockchains, such as Bitcoin or Ethereum, are open and permissionless networks that allow anyone to join and participate in the ledger maintenance and validation, which consumes more energy but provides more transparency and decentralization [12]. Private blockchains, such as Hyperledger Fabric or Corda, are closed and permissioned networks that allow only authorized entities to join and participate in the ledger maintenance and validation, which consumes less energy but provides less transparency and decentralization

The protocols used to communicate the blockchain peers also have a relevant impact on energy consumption. For instance, many blockchains rely on flooding techniques to propagate the blocks, which results in duplicates and in an inefficient use of the existing bandwidth \cite{Antwi2022}. In addition, blockchains can make their peers select randomly with whom they exchange transaction data, thus limiting potential throughput increases. As a consequence, to minimize the overall blockchain energy consumption, network communications should be analyzed and optimized accordingly \cite{Antwi2022}.

\subsection{Data Storage and Validation}

While data storage usually does not use much energy, the validation and verification of such data within a blockchain system consumes a significant amount of computational resources and energy \cite{Zysk2015}. Specifically, the continuous cryptographic operations that need to be performed involve a high energy consumption, but they are required to provide security and integrity to the data and transactions. Specifically, different cryptographic operations have different energy requirements and trade-offs. For example, hash functions such as SHA-256 or Keccak-256 are used to generate unique identifiers for blocks and transactions, which consume a moderate amount of energy but provide high security and collision resistance. Digital signatures like ECDSA can be used to verify the authenticity and ownership of transactions, which consume a low amount of energy but require public-key infrastructure and certificate authorities.


\subsection{Mining}

\begin{figure*}[!hbt]
     \centering
     \includegraphics[scale=0.4]{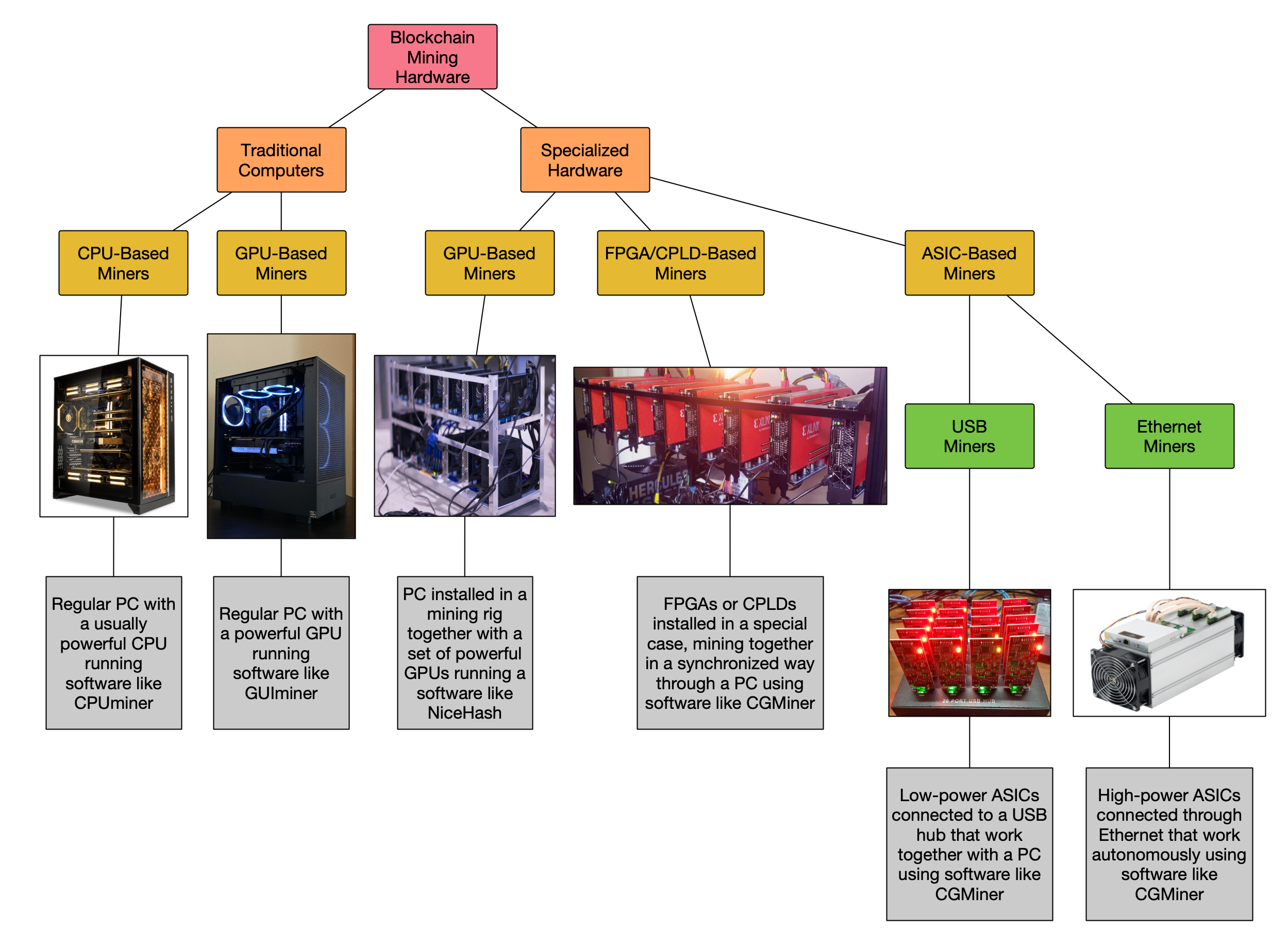}
     \caption{Types of hardware for blockchain mining.}
     \label{figure:miningHardware}
\end{figure*}

As it was previously described in Section \ref{sec:overview}, mining activities, especially in blockchains that use PoW, use a lot of energy. The process of finding solutions to complex cryptographic puzzles requires a lot of computational power, leading to increased energy consumption. The energy consumption of mining depends on factors such as the mining hardware used, the difficulty of the puzzles, and the energy source powering the mining operations \cite{DeVries2018}. The main types of blockchain miners/validators are illustrated in Figure \ref{figure:miningHardware} and include:

    \begin{itemize}
        \item Miners based on traditional computers. They use regular computers (usually with powerful CPUs and/or GPUs) that run blockchain software to carry out the required mining/validation operations. This kind of hardware is really flexible, but such a flexibility comes at the cost of being able to perform less operations per second than dedicated hardware.

        \item Specialized hardware. Such hardware makes use of dedicated hardware like FPGAs, CPLDs or ASICs. FPGAs and CPLDs are in general less powerful than ASICs, but they have the benefit of being able to be reprogrammed. ASICs cannot be reprogrammed, but they are currently the most powerful hardware for blockchain mining/validation.
    \end{itemize}

\begin{figure*}[!hbt]
     \centering
     \includegraphics[scale=0.22]{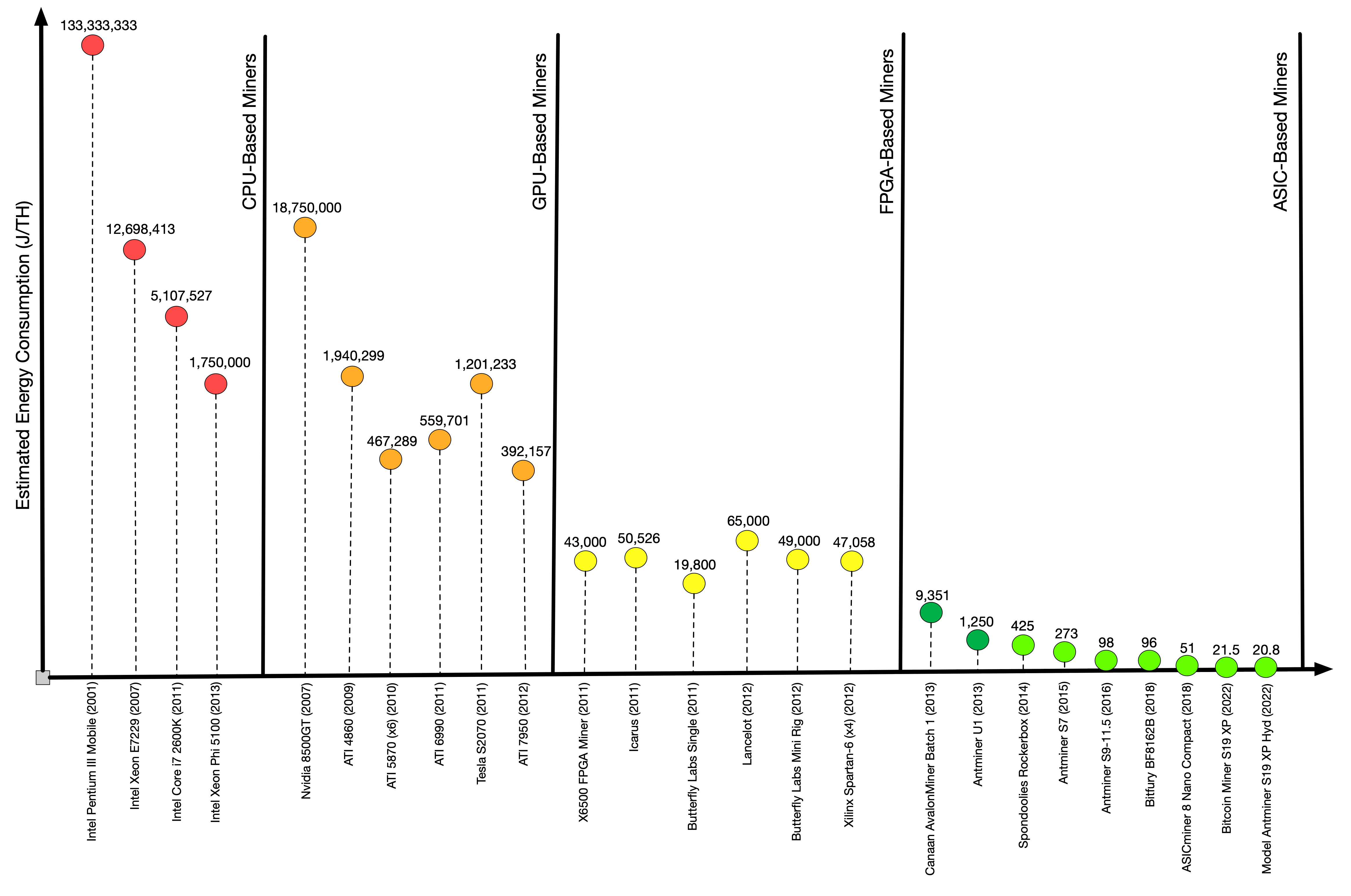}
     \caption{Energy consumption of different miners for SHA-256.}
     \label{figure:miningEnergy}
\end{figure*}

    Figure \ref{figure:miningEnergy}, whose data comes from \cite{Sinan2019,Romano2017,Cocco2016,Amila2019,S19XP}, compares the energy consumption (in terms of Joules per TeraHash) for SHA-256 miners based on CPUs, GPUs, FPGAs and ASICS. As it can be observed, ASIC-based mining is currently clearly the most energy efficient, while CPU-based miners are the ones that require more power, being GPU and FPGA-miners in the middle (however, note that it is usual to build GPU and FPGA mining clusters, which jointly decrease energy consumption). Moreover, mining has become more sustainable through time: the latest miners are several orders of magnitude more efficient than the ones that existed ten years before.

\subsection{Block size}

The size of the blocks of a blockchain determines the number of transactions that are packed and processed together during mining. Although the energy consumption related to the mining process should remain approximately constant (i.e., the amount of energy per transaction should be almost the same, independently of the block size), it is true that block size impacts other processes required by a blockchain \cite{Sedlmeir2020}: the larger the block size, the longer it takes to propagate it through the network, which increases blockchain latency (i.e., the time required to deliver a block to the nodes), it requires more time and resources for validation/mining, and more storage capacity is necessary. In addition, especially in PoW-based blockchains, the longer the block size, the more powerful mining hardware needs to be, thus increasing the overall mining energy consumption. 

Moreover, it should be noted that increasing block size excessively has a negative impact on the blockchain security: if powerful mining hardware is necessary, many low computational-resource, low-bandwidth and low-storage capacity miners would be excluded from the mining process, thus making the network more prone to 51\% attacks (i.e., the network could be controlled more easily by a pool of powerful miners). In contrast, the lower the block size, the lower the resources required to participate in the mining process, so the cheaper is to create rogue nodes for Sybil attacks.
Therefore, a trade-off should be achieved among block size, mining energy consumption and security.

\section{Strategies to Create Green Blockchains}
\label{sec:strategies}

\subsection{Consensus Mechanism}

\begin{itemize}
    \item PoW. While PoW has proven to be secure, it is highly energy-intensive due to the computational power required to mine blocks \cite{DeVries2018}.  Nonetheless, researchers have already proposed green-PoW consensus algorithms. For instance, in \cite{Lasla2022} the authors suggest using a dynamic difficulty adjustment scheme to reduce the energy consumption of PoW mining. The scheme adjusts the difficulty level of the puzzle based on the number of active miners and their hash rates. The authors showed that their green-PoW consensus mechanism can reduce energy consumption up to 50\% without compromising security or performance.

    \item PoS. As it was previously described in Section \ref{sec:consensusMechanisms}, PoS selects validators to create new blocks based on their stake or ownership of the tokens or cryptocurrency \cite{Peercoin}. Thus, PoS eliminates the need for solving computationally expensive problems like PoW, so it significantly reduces energy consumption (several orders of magnitude with respect to PoW), as no computational work is involved in block creation. Moreover, in general, PoS performance and energy consumption does not depend on the network size, hence being really efficient when used for large-scale systems \cite{Sedlmeir2020}. Examples of blockchains that are already using PoS or one of its variants are EOSIO and Ethereum. For instance, thanks to migrating Ethereum from PoW to PoS, it is estimated that energy consumption decreased 99\% \cite{Heinonen2022}. However, it should be noted that some authors indicate that avoiding the use of PoW reduces security (the blockchain control might end up being controlled by a small elite that holds most of the `capital'), so it requires careful design to ensure security while reducing energy consumption \cite{Kiayias2017,Buterin2017}. 

    \item DPoS. As mentioned in Section \ref{sec:consensusMechanisms}, DPoS introduces a small group of elected nodes as delegates who are responsible for validating transactions and producing blocks \cite{Larimer2014}. It achieves faster transaction speeds and lower energy consumption compared to PoW \cite{Platt2022}. However, while DPoS provides higher performance and energy efficiency than PoW and PoS, it requires active participation from the community to maintain decentralization and avoid potential centralization risks. 

    \item BFT. BFT relies on communications and agreement between nodes, and since BFT algorithms often reach consensus quickly (usually in a matter of seconds), it does not require heavy computational resources and, as a consequence, it consumes much less energy than other consensus protocols \cite{Vuko2016}. It must be indicated that, while BFT excels in permissioned and small-scale networks, it may face scalability issues in fully decentralized networks with many nodes due to communications overhead (i.e., since  BFT requires nodes to communicate extensively with each other, the protocol can become a bottleneck in large highly-decentralized networks). Nonetheless, some authors have recently proposed optimizations to mitigate this problem \cite{Oh2024}.

    \item PoA. PoA is especially useful in industries and for the public sector, where the involved participants group into a consortium where all of them are known. In such a case, where there is a certain degree of trust on the participants, is not necessary to consume a precious resource like energy to protect the system, so voting and validation can be carried out more efficiently (e.g., by selecting a random trusted validator or by performing a poll among the participants where each participant has a vote).


    \item PoB. PoB is more energy efficient than PoW-based consensus mechanisms, but it must be noted that, since it requires nodes to purchase tokens periodically, it can derive into an indirect energy use that does not occur in other consensus protocols like PoS. Therefore, the developers of PoB-based blockchains need to select appropriate network parameters to minimize token burning while maintaining the performance of the network.

    \item PoC. Proof-of-Capacity involves mining: when a new block needs to be mined, the network broadcasts a challenge and then miners scan their precomputed plots stored on their hard drives to find the closest match to the solution. The miner whose stored data contains the closest hash to the challenge wins the right to create the next block and claim the reward. Therefore, the mining process is more about searching through pre-plotted data rather than generating new computations, which makes PoC less energy-intensive than PoW. Nonetheless, it worth noting that PoC has been critized in terms of sustainability, since participants may buy large numbers of hard drives specifically for mining and then discard them when they become obsolete or unprofitable. 

    \item PoAC. Proof-of-activity, which implements a PoW-based and a PoS-based phase, is more energy efficient than traditional PoW (since PoW mining work is reduced to the initial phase), but it consumes more energy than PoS. As a consequence, developers should try to minimize the impact of the PoW-based phase in order to make the overall PoAC consumption as close as possible to the one produced by the equivalent PoS-based system.

    \item Other consensus mechanisms can be used at the same time for the regular operation of the blockchain and also for other purposes, so their energy consumption can be somehow justified. For instance, Proof of Solution (PoSo) replaces meaningless PoW puzzles with a meaningful optimization problem. Another example is  Proof-of-Play (PoP), which harnesses the popularity of online games for helping with the consensus \cite{Yuen2019}. Specifically, in PoP the mere act of playing is enough for contributing to mining blocks, as it has been demonstrated in blockchains like Motocoin \cite{Motocoin} or Huntercoin \cite{Huntercoin}.

\end{itemize}

\subsection{Network Architecture}


A good survey on blockchain network optimization techniques can be found in \cite{Antwi2022}, which includes, among others, the following strategies:

\begin{itemize} \item Optimization of peer-to-peer (P2P) network protocols. Optimizing P2P network protocols can help lower energy consumption in blockchain systems. Some strategies for network protocol optimization include:

\begin{itemize}
    \item Routing Efficiency. Enhancing the efficiency of routing algorithms and protocols can reduce the energy needed for message propagation and decrease network overhead \cite{Antwi2022}. For instance, some researchers have found, after analyzing thoroughly Ethereum gossip protocol (i.e., the mechanism used to propagate blocks), that only a small amount of peers are useful during the propagation of new blocks and that the physical location clearly affects when nodes hear about new blocks \cite{Kiffer2021}.

    \item Peer Discovery. Efficient peer discovery mechanisms help nodes join the network faster, reducing the energy spent in searching for and establishing connections. In fact, it is recommended that peers select neighbors that are physically close to reduce network delay \cite{Nguyen2019}.

    \item Network Topology. Designing an optimized network topology, such as using overlay networks or hierarchical structures, can improve scalability and reduce energy consumption by minimizing the distance and number of hops required for data propagation \cite{Sedlmeir2020}. In fact, one of the strategies to reduce energy consumption is to reduce the degree of redundancy in network topologies: the smaller the number of nodes that are needed to perform certain operations, the smaller the overall energy consumption. In this aspect, researchers studied the impact of logical and physical networks on their performance and found a significant degree of traffic redundancy \cite{Zhang2021}. To avoid such problems, there are techniques like sharding \cite{Yu2020}, which allows for creating subsets of nodes called `shards' that are responsible for processing certain transactions (i.e., instead of using the whole blockchain nodes for mining/validation, only a subset of nodes participate in the processing of each transaction). 
    Thus, sharding acts like a partitioning technique that allows for splitting a large database into smaller parts that are distributed among the members of a shard \cite{Baniata2023}, making large database access faster and its data more manageable. However, despite the benefits of sharding, it needs to be adapted to the used consensus mechanism. For example, it is difficult to create efficient PoW-based blockchains, since the estimated computing power used by a shard needs to be balanced among all the nodes. In contrast, in a PoS-based blockchain, since the stake of each node is known publicly, it is easier to create well-balanced shards. As a consequence, green sharding mechanisms need to consider decentralization, scalability, security and energy efficiency. More details about sharding and a comparison among state-of-the-art sharding solutions can be found in \cite{Baniata2023}.


\end{itemize}

\item Reduction of network latency and bandwidth requirements. Some approaches to achieve such a reduction consist in the use of:

\begin{itemize}
    \item Compression techniques. Applying compression techniques to network data can reduce the amount of transmitted data, thus reducing network latency and bandwidth requirements \cite{Antwi2022}.
    
    \item Caching and Content Delivery Networks (CDNs). Using caching mechanisms and CDNs can store and serve frequently accessed data closer to the users, reducing the need for data retrieval from distant network nodes and minimizing network latency.
    
    \item Efficient protocol design. The development of lightweight and efficient network protocols adapted to blockchain systems can reduce the amount of exchanged data, decrease network latency and improve overall energy efficiency \cite{Antwi2022}.
\end{itemize}

\item Use of off-chain solutions to minimize on-chain transactions. Off-chain solutions can reduce the number of transactions that need to be processed and validated on the blockchain, therefore reducing the energy consumption associated with consensus and mining. Off-chain solutions include:

\begin{itemize}
    \item Sidechains and State Channels. They allow transactions to be executed on separate chains or channels that are linked to the main blockchain, reducing the load on the main chain and increasing transaction throughput. Plasma, for Ethereum, is an example of sidechain technology that allows for creating small blockchains that report periodically to the main network.
    
    \item Layer 2 Protocols. Layer 2 protocols are solutions that operate on top of the blockchain layer, providing scalability and efficiency improvements without compromising security or decentralization. Examples of layer 2 protocols include Lightning Network and Raiden Network \cite{Antwi2022}. Thus, the problem is essentially the low transaction throughput of certain Layer 1 protocols. For example, in Bitcoin, before the release of SegWit (before the SegWit update, Bitcoin blocks were limited to a size of roughly one megabyte) and of the Lightning Network, it was only possible to perform roughly seven transactions per second. The Lightning Network allows for performing thousands of transactions per second without increasing the energy consumption related to mining \cite{Heinonen2022}. The Raiden Network is similar to the Lightning Network but for Ethereum: it provides a fast and cheap micropayment channel that records transactions first off-chain and then on-chain.
    
\end{itemize}
\end{itemize}

\subsection{Data Storage}

The process of storing and validating the information exchanged in a blockchain also consumes a significant amount of energy. To minimize such consumption, future developers and researchers should consider the use of:

\begin{itemize} 
    \item Compression techniques for blockchain data. Applying compression techniques for blockchain data can significantly reduce the energy consumption related to data storage and transmission. Compression techniques aim to reduce the size of data without compromising its integrity. Some strategies for data compression in blockchain systems include:

    \begin{itemize}
        \item Lossless compression. Lossless compression algorithms, such as gzip or deflate, reduce data size without losing any information. This can help to decrease storage requirements and network bandwidth usage, leading to energy savings. 
    
        \item Transaction aggregation. Aggregating multiple transactions into a single compressed data structure can reduce the overall data size \cite{Adaptive2020}. By bundling transactions together, the number of data operations and storage requirements can be minimized. 
    
        \item Merkle trees. Merkle trees are hash trees that allow for efficient verification of data integrity \cite{Nakamoto2008}. By representing a large amount of data with a compact hash, Merkle trees reduce the storage and computational overhead required for data validation. 
        
    \end{itemize}

    \item Pruning and archiving of unnecessary data. Pruning and archiving strategies help to remove or to store unnecessary or outdated data, reducing the energy consumption related to data storage. Specifically:
    
    \begin{itemize}
        \item Pruning involves removing unnecessary data from the blockchain, such as spent transaction outputs or older transaction history that is no longer required for validation \cite{Selective2018}. Since pruning reduces the storage requirements, it can lead to energy savings.
    
        \item Archiving involves moving data that are not frequently accessed or required for validation to secondary storage or off-chain storage solutions. By storing these data outside the main blockchain, energy consumption can be reduced.
    \end{itemize}

    \item Decentralized storage. To create a complete decentralized blockchain-based solution, data should be also distributed and processing should be decentralized \cite{Fernandez2019}. Such decentralization provides redundancy and security to prevent \ac{DoS} attacks. However, it must be noted that decentralized storage requires storage nodes to be synchronized periodically, which increases the energy consumption dedicated to peer communications.

\end{itemize}

\subsection{Data validation}

As it was previously mentioned, data validation is essentially related to  cryptographic operations, which are another energy-intensive component of blockchain systems, as they provide security and integrity for the data and transactions on the ledger. Different cryptographic operations have different energy requirements and trade-offs. 
In fact, some authors have already analyzed mining hardware power consumption and determined that the hashing algorithm mainly determines the mining efficiency \cite{Li2019}. 
For instance, in \cite{Bozi2022} the authors propose optimized threshold implementations for securing cryptographic accelerators for low-energy and low-latency applications. Threshold implementations are a masking countermeasure that can protect cryptographic operations from side-channel attacks. The authors proposed three optimization techniques that can reduce the number of output shares, the number of non-linear gates, and the number of random bits required for threshold implementations. The authors showed that their techniques can achieve significant energy savings and latency reduction for various cryptographic primitives, such as AES, Keccak or SHA-3. 

Currently, many blockchains required all selected miners/validators to carry out all the steps necessary to add a block to the blockchain, thus needing to perform all the involved cryptographic operations. A strategy to reduce energy consumption would consist in only performing short data correctness proofs. This is what is proposed by Zero-Knowledge Proofs (ZKPs) like SNARKS or STARKS, which require much less computation and communications overheard than traditional blockchain cryptographic verification mechanisms \cite{Khor2022}.



\subsection{Smart Contract Execution}

Smart contracts are also essential for some of the most advanced blockchains, so their execution efficiency has a significant impact on the overall consumption of the network. To reduce such consumption, developers should consider the following main factors:

\begin{itemize} 
    
    \item Optimization of smart contract code and execution. Optimizing smart contract code and execution can help to reduce the energy consumption of smart contracts. Some strategies for optimization include:

    \begin{itemize}
        \item Gas optimization. Gas is a unit used by Ethereum to measure computational effort in smart contracts. Optimizing the code to reduce gas consumption can lead to energy savings. Techniques such as minimizing storage operations, using efficient algorithms, and avoiding unnecessary computations can help to optimize gas usage \cite{Khan2021}.
    
        \item Loop and recursion efficiency. Efficient utilization of loops and recursion in smart contract code is essential. Reducing the number of iterations or implementing efficient loop and recursion patterns can minimize the computational workload and energy consumption \cite{Vacca2021}.
    
        \item Gas limit estimation. Accurately estimating the gas limit required for smart contract execution helps to prevent unnecessary gas wastage. Adequate estimation ensures that the contract is executed within the available resources, reducing the energy consumed by excessive gas usage \cite{Khan2021}.
    \end{itemize}

    \item Integration of energy-efficient programming languages. Integrating energy-efficient programming languages can help to reduce the energy consumption of smart contract execution. Some programming languages are designed to be more energy-efficient or provide features that optimize resource usage. Considerations include:

    \begin{itemize}
        \item Low-level languages: Low-level languages like Rust or C++ provide more control over resource usage and allow for fine-grained optimizations, potentially leading to more energy-efficient execution \cite{Ranking2021}. 
    
        \item High-level languages: High-level languages like Solidity (used in Ethereum) or Vyper offer built-in gas optimization features and higher-level abstractions, enabling developers to write more concise and readable code, potentially leading to more energy-efficient executions \cite{Vacca2021}. 
        
    \end{itemize}
    
\end{itemize}

\subsection{Mining and Block Creation}

As it has been previously described, mining can derive into a significant portion of the global consumption of a blockchain network. As a consequence, green developers should consider the following aspects:

\begin{itemize} 
    \item Hardware optimization for mining operations. Optimizing hardware for mining operations can help reduce the energy consumption of blockchain systems. Here are some strategies for hardware optimization:

    \begin{itemize}
        \item Energy-efficient mining equipment. Using energy-efficient mining equipment such as ASICs or FPGAs can improve the efficiency of the mining process. These specialized devices are designed to perform mining computations more efficiently, consuming less energy per hash calculation. 
        Alternatively, some authors propose to make use of photonic miners to reduce energy consumption \cite{Dubro2020}.
    
        \item Cooling and power management. Implementing efficient cooling systems and power management techniques for mining equipment can reduce energy waste. Proper cooling prevents overheating and ensures optimal performance, while power management techniques minimize energy usage during idle or low-load periods.
    
        \item Hardware upgrades and maintenance. Regular hardware upgrades and maintenance help ensure optimal performance and energy efficiency. Upgrading to more energy-efficient hardware or replacing faulty components can help to reduce energy consumption during mining operations. 
    \end{itemize}

    \item Transition to renewable energy sources for mining activities. Transitioning mining activities to renewable energy sources is essential for reducing the carbon footprint of blockchain systems. Some approaches for integrating renewable energy include:

    \begin{itemize}
        \item Solar and wind power. Installing solar panels and wind turbines to power mining operations can utilize clean and renewable sources of energy. Locating mining facilities in areas with abundant sunlight or strong winds can optimize the use of these renewable resources. There are specific blockchains designed to incentivize the use of solar energy, like the one used by SolarCoin \cite{SolarCoin}.

        \item Hydropower. Utilizing hydropower, which harnesses the kinetic energy of flowing water, to power mining operations can provide a reliable and sustainable source of energy. Locating mining facilities near rivers or dams can optimize the use of hydropower. In fact, some reports have indicated that in 2019 74\% of the Bitcoin mining operations relied heavily on renewable energy sources due essentially to the availability of hydropower \cite{Kshetri2022}. Such a percentage decrease significantly in the last years (to roughly 25\%) after China outlawed cryptomining, which made miners move from mountainous Chinese regions (where hydropower was prevalent) to the US (where gas provides much of the generated power).

        \item Geothermal power. Some countries like Iceland can take advantage of cheap electricity from geothermal plants to power blockchain equipment and infrastructure \cite{Hammons2008}.

    \end{itemize}

    Nonetheless, it must be noted that some environmentalists have complained on the excessive energy use of mining plants and on the use of certain renewable resources, specifically in relation to the use of water for cooling \cite{Kshetri2022}. In fact, some researchers have concluded that the sole use of renewable energy sources is not the answer for the sustainability problem related to some blockchains \cite{Vries2019}.

    \item Reutilization of the mining results to serve practical purposes and thus justify the energy investment. Such a reutilization has been previously related to PoW inefficiency and to what was coined as 'bread pudding protocols' \cite{Jakobson1999}.
    For instance, researchers have proposed to train deep learning models \cite{Chenli2019}, to execute genetic algorithms in a collaborative way \cite{Bizarro2020}, to look for prime numbers \cite{Primecoin} or to contribute to scientific research \cite{Gridcoin,Foldingcoin}. 
    
    \item Hash reuse. Part of the results of mining is the computation of many hashes, which may be reused to avoid its recalculation, thus coining the term `hash recycling' \cite{Heinonen2022}. This has already been performed for other security scenarios through Hellman Tables\cite{Hellman1980}, Rainbow Tables \cite{Oechslin2003} or one of their variants \cite{Kara2009}.

    \item Merge mining. It allows to mine the blocks of two or more blockchains at the same time. Thus, the energy consumption related to the effort put on the consensus protocol can be shared among multiple blockchains. Nonetheless, since merge mining can end up centralizing mining, more effort should be put in its impact \cite{Judmayer2017}.

\end{itemize}

\subsection{Network optimization techniques}

In blockchain networks, network optimization involves methods that enhance performance by accelerating the communications among nodes, by lowering bandwidth usage or by implementing strategies that maintain efficient performance for large-scale applications. Thus, current network optimization techniques in blockchain focus essentially on improving four aspects:

\begin{itemize}
    \item Enhancing P2P. Some authors have already evaluated the efficiency of the P2P protocols implemented in certain blockchain networks in order to analyze their optimization \cite{Shaleva2021}. Specifically, some researchers targetted the Bitcoin P2P protocol \cite{Vu2019}, while others have proposed generic optimizations aimed at accelerating transaction speed thanks to improvements in the P2P layer (in contrast to the improvements related to other layers, like the ones implemented through Sharding, State channels or Plasma) \cite{Yang2019}. Moreover, other authors, after observing that P2P topology impacts broadcast speed of blockchain data significantly (what leads to poor performance and hence unnecessary energy consumption), propose a protocol based on fast and scalable broadcasts \cite{Hao2019}.

    \item Reducing duplicate messages caused by gossiping. Some researchers suggested improving blockchain gossip algorithms \cite{He2019}, while others ease its scalability by defining transmission paths and neighbor node subareas \cite{Yu2021}. Other authors focused on specific blockchains and studied how to make their gossip-based protocols more efficient \cite{Shaleva2021-B}.

    \item Minimizing the size of the data exchanged between blockchain nodes. Some authors have proposed packet aggregation schemes \cite{Ahn2020}, while others proposed speeding up block propagation by exploiting rateless erasure codes \cite{Zhang2022}.

    \item Reducing the communications complexity of the consensus mechanisms, thus enabling faster interactions, reducing congestion risks and minimizing bandwidth consumption \cite{Jin2019,Zhao2021}. Other researchers were aimed at accelerating message propagation \cite{Kan2018,Santiago2020,Huang2021}.

\end{itemize}

\section{Challenges and Limitations of implementing Green Blockchains}
\label{sec:challenges}

The implementation of energy-efficient solutions for blockchain systems is not a trivial task, as it involves various challenges and limitations that need to be addressed. The next subsections describe the main challenges and limitations for such an implementation.

\subsection{Potential trade-offs}

 Energy-efficient solutions may introduce trade-offs between energy consumption and other performance or security metrics, such as throughput, latency, scalability, reliability or resilience. For example, reducing the number of nodes or the frequency of communications in the network may reduce energy consumption, but they also reduce the throughput or reliability of the network. Similarly, using simpler or fewer cryptographic operations may reduce energy consumption, but also reduce the security or integrity of the data or transactions. Therefore, finding the optimal balance between energy consumption and other metrics is a challenging task that requires careful analysis and evaluation.

In addition, achieving energy efficiency in blockchain systems can sometimes come at the expense of security and robustness, so it is important to consider the potential trade-offs that may arise. Some challenges include:

\begin{itemize}
    \item Consensus security. Energy-efficient consensus mechanisms such as PoS may introduce new security vulnerabilities compared to traditional PoW mechanisms. It is crucial to carefully design and evaluate the security implications of energy-efficient consensus protocols to ensure the system remains resilient against attacks.

    \item Decentralization. Energy reduction strategies such as offloading computation or consolidating mining pools may impact the decentralization of the blockchain network. For instance, researchers that monitored Ethereum during multiple months detected that the top 15 miners were responsible for mining over 90\% of the blocks and that the blocks mined by 3 top mining pools had much greater likelihood of been included in the blockchain \cite{Kiffer2021}. Moreover, centralization can potentially introduce single points of failure or increase the risk of collision among a smaller number of participants. Balancing energy efficiency with the goal of maintaining a decentralized network is a complex challenge. This is something that needs to be considered with care, specially in permissioned networks: in an extreme case only one validator node would exist, thus creating a de-facto centralized network. As a consequence, permissioned blockchains, which in practice may be energy efficient need to pay close attention to the entry barriers they impose on the blockchain participants.

    
\end{itemize}

\subsection{Energy consumption estimation}

The estimation of the energy consumption of blockchain systems is not straightforward due to their multiple components, the available configuration parameters and the way nodes interact.
Nonetheless, in the last years two main approaches have been followed for estimating the energy consumption of a blockchain \cite{Platt2022}:

\begin{itemize}
    \item To quantify the energy consumption of a representative participant and then extrapolate such a measurement to the rest of the network. For example, in \cite{Li2019} the authors measure the consumption of mining hardware and then estimate the power consumption of the whole blockchain.

    \item To create a mathematical model that estimates energy consumption by considering the essential metrics of the blockchain. Many of these estimations rely of publicly available data (e.g., the hash rate of a blockchain) and relates them with specific hardware \cite{Zade2019,Galler2020,Lei2021}. For instance, in \cite{Platt2022} the authors propose a simple energy consumption model that can be applied to blockchains that make use of PoS consensus mechanisms. Such a model estimates the energy consumption per transaction by considering factors like the number of validators or the network throughput. However, it is worth pointing out that this kind of models are created to avoid performing time-consuming experimental validations, thus sacrificing precision \cite{Platt2022}. Moreover, researchers need to be aware that some publications make use of energy consumption figures that do not come from controlled experiments, but from `promotional materials', so the extracted conclusions may be skewed or simply wrong \cite{Platt2022}.

\end{itemize}

The literature also contains examples of works from researchers that focused on the best practices for determining blockchain energy consumption \cite{Lei2021}, while others analyzed the state of the art on blockchain energy-consumption estimation, focusing on PoW-based systems and determining lower and upper bounds \cite{Sedlmeir2020}. Nonetheless, such authors justify the validity of the proposed upper bound arguing that the energy consumption related to maintaining the network (i.e., for blockchain download and initialization, and for peer communications) are negligible, which may be true for certain PoW-based blockchains, but which is not necessarily true for some of the latest blockchains.

Furthermore, some authors have estimated and compared the consumption of PoW and PoS-based blockchain systems with traditional financial transaction processing systems like VisaNet payment network \cite{Platt2022}, concluding that:

\begin{itemize}
    \item PoW-based systems like Bitcoin was at least three orders of magnitude higher than the highest consuming PoS-based system they evaluated.

    \item There are already PoS-based systems that, in certain configurations, are able to consume less energy per transaction and globally than VisaNet. Nonetheless, the authors of \cite{Platt2022} admit that they are not aware of any PoS-based system able to reach the throughput levels of VisaNet, although promising alternatives are being studied and deployed (e.g., the Lightning network, optimistic rollups o ZK-based rollups).

\end{itemize}

Nonetheless, for the sake of fairness, it is worth pointing out that some reports suggest that the Bitcoin network consumes less than a half of the energy required by the large data centers used by traditional banks \cite{Kshetri2022} and that, in fact, mining is greener than it is generally expected \cite{Heinonen2022}.

\subsection{Regulatory and governance challenges in implementing energy-saving measures}

Implementing energy-saving measures in blockchain systems can face regulatory and governance challenges like:

\begin{itemize}
    \item Regulatory compliance. Blockchain projects operating in different jurisdictions may face varying regulatory frameworks concerning energy consumption, renewable energy sources and environmental sustainability. Adhering to relevant regulations and compliance requirements adds complexity to the implementation of energy-saving measures.

    \item Coordination among stakeholders. Implementing energy-saving measures often requires collaboration among various stakeholders, including blockchain developers, miners, users, energy providers and regulatory bodies. Achieving consensus and coordination among these stakeholders can be challenging, especially when conflicting interests or incentives exist.
    
\end{itemize}

\subsection{Scalability and performance implications of energy reduction strategies}

Energy reduction strategies must also consider the scalability and performance implications they may introduce. Key considerations include:

\begin{itemize}
    \item Scalability. Implementing energy-saving measures should not compromise the scalability of the blockchain system. As the network grows and transaction volumes increase, energy-efficient mechanisms should be able to handle the load and ensure efficient transaction processing.

    \item Performance impact. Energy reduction strategies should be evaluated for their potential impact on the performance of the blockchain system. For example, offloading computations to external platforms may introduce additional latency or dependencies on third-party services, which can affect the overall performance and user experience.

    \item Cost efficiency. While reducing energy consumption is a primary goal, it is also important to consider the cost efficiency of implementing energy reduction strategies. Solutions that reduce energy consumption but introduce significantly higher operational costs may not be sustainable in the long term.
\end{itemize}

\subsection{Other challenges}

Besides the previously mentioned challenges, the following aspects should also be considered by future green blockchain researchers:

\begin{itemize}
    \item Some energy-efficient solutions may have limited experimental validation or empirical evidence to support their claims or assumptions. For instance, the use of reversible computing has been proposed for Bitcoin mining, but it is currently not known with precision how much energy would be saved in comparison with traditional ASIC-based mining \cite{Heinonen2022}. Ternary computing has also been suggested for reducing DLT energy consumption (e.g., by IOTA, but the limited commercial hardware support has not allowed its validation at a massive scale \cite{Heinonen2022}).

    \item Some energy-efficient solutions may have limited compatibility or interoperability with existing standards or protocols. 

    \item Besides all the previously mentioned technical challenges, there is a need for education on the use of green blockchain and DLTs. A risk exists on the fact that greener technologies can spread its use at a massive scale, thus increasing the overall consumption. This effect has already been observed regarding the use of LED lighting: the improvements on energy efficiency derived into using more LEDs and hence on consuming more light \cite{Hicks2015}.

\end{itemize}

\section{Conclusion}
\label{sec:conclusions}

Blockchain  has been regarded in the past as an energy-inefficient technology essentially to the prejudices that arose together with the popularization of Bitcoin, whose PoW consensus mechanism was actually power hungry.
However, since the inception of Bitcoin in 2008, blockchain technologies have evolved significantly and many authors have already proposed diverse strategies to create Green Blockchains.
Thus, this article reviewed and analyzed such strategies with the objective of reducing the energy consumption of the main energy-intensive components of a blockchain system. 
For such a purpose, after discussing the background work and the importance of addressing energy consumption in blockchain systems, the main blockchain components were analyzed, including consensus mechanisms, network architectures, data storage and validation, smart contract execution, or mining and block creation. Then, multiple useful strategies to improve the energy efficiency of such blockchain components were detailed. Moreover, the most relevant challenges and limitations of implementing energy-efficient blockchain-based solutions have been described.
As a consequence, this article provides precise insights and guidance to future researchers for the development of the next generation of Green Blockchains.

\section*{Data availability}
No datasets were generated or analyzed during the study. 

\section*{Acknowledgements}
This work has been funded by grant TED2021-129433A-C22 (HELENE) funded by MCIN/AEI/10.13039/501100011033 and the European Union NextGenerationEU/PRTR. 

\section*{Author contributions statement}
Conceptualization, T.M.F.-C.; methodology, T.M.F.-C. and P.F.-L.; investigation, T.M.F.-C. and P.F.-L.; writing—original draft preparation, T.M.F.-C. and P.F.-L.; writing—review and editing, T.M.F.-C. and P.F.-L.; supervision, T.M.F.-C.; project administration, T.M.F.-C.; funding acquisition, T.M.F.-C. 



 
\section*{Competing interests}
The authors declare no competing interests.

\end{document}